%% file: main.tex
\documentclass[sigplan,twocolumn]{acmart}
\renewcommand\footnotetextcopyrightpermission[1]{}
\input{preamble}

\begin{document}
\input{macros}
\pagestyle{plain}

\title[]{\name: Exploiting Communication Sparsity for Sequence-Parallel Video DiTs}

\author{Desen Sun}
\affiliation{%
  \institution{University of Waterloo}
  \city{Waterloo}
  \country{Canada}
}
\email{desen.sun@uwaterloo.ca}

\author{Xinrui Zhong}
\affiliation{%
  \institution{Rice University}
  \city{Houston}
  \country{USA}
}
\email{xz118@rice.edu}

\author{Yuke Wang}
\affiliation{%
  \institution{Rice University}
  \city{Houston}
  \country{USA}
}
\email{yuke.wang@rice.edu}

\author{Sihang Liu}
\affiliation{%
  \institution{University of Waterloo}
  \city{Waterloo}
  \country{Canada}
}
\email{sihangliu@uwaterloo.ca}

\input{sec/0_abs}


\maketitle

\input{sec/1_intro}
\input{sec/2_back}
\input{sec/3_motiv}

\input{sec/4_selectkv}

\input{sec/5_ring}
\input{sec/6_ce}
\input{sec/7_chunk}

\input{sec/8_eval}

\input{sec/9_con}

\bibliographystyle{ACM-Reference-Format}
\bibliography{bib/sys,bib/misc,bib/ml}
\end{document}

%% file: preamble.tex
\usepackage{subcaption}
\usepackage{xspace}
\usepackage{multicol}
\usepackage{multirow}
\usepackage{booktabs}
\usepackage{cleveref}
\usepackage{graphicx}
\usepackage[super]{nth}
\usepackage{enumitem}
\usepackage{svg}
\usepackage[normalem]{ulem}
\usepackage{url}
\usepackage{amsmath}
\usepackage[skins]{tcolorbox}

\newcounter{takeawaycounter}
\crefname{takeawaycounter}{Take away}{Take aways}

\newtcolorbox[use counter=takeawaycounter]{takeaway}[2]{
  enhanced, width=\hsize,left=1pt,right=1pt,top=1pt,bottom=1pt,colback=gray!6!white,boxrule=1pt,colframe=gray!30!white,
  label type=takeawaycounter,
  label = #1,
  title = \textcolor{black}{\bf Take away~\arabic{takeawaycounter}~#2}
}

\usepackage{pifont}
\newcommand{\circled}[1]{\scalebox{1.25}{\ding{\numexpr#1+201}}}

%% file: macros.tex
\newcommand{\name}{\textsc{SparSP}\space}
\newcommand{\figplaceholder}[2][1.15in]{%
  \fbox{\parbox[c][#1][c]{0.94\linewidth}{%
    \centering\small\textsf{Figure placeholder}\\[0.4em]#2}}}

\newcommand{\sihang}[1]{\textcolor{red}{SL: #1}}

\newcommand{\desen}[1]{\textcolor{blue}{desen: #1}}

\newcommand{\erjing}[1]{\textcolor{orange}{EL: #1}}

%% file: sec/0_abs.tex
\begin{abstract}
Diffusion Transformers have become the dominant architecture for video generation. Their substantial computational cost motivates scaling inference across multi-GPU servers, yet efficient scaling remains challenging on commodity GPUs connected via PCIe, whose bandwidth is limited. Although sparse attention substantially reduces computation, its implications for communication remain underexplored.

This paper argues that attention sparsity should be treated as a communication primitive. We present \name, an efficient sparse sequence parallel communication system that co-designs token distribution, communication routing, and asynchronous execution for sparse video diffusion models. First, Dependency-Aware Placement distributes sequence blocks according to diffusion models' sparse attention patterns. Second, Demand-Directed KV Routing transfers KV blocks directly to requesting GPUs without intermediate relays. Third, a Decoupled Transfer Runtime separates communication from GPU computation to reduce resource contention and maximize effective bandwidth. Our evaluation shows that \name improves attention performance by 1.38--1.5$\times$, achieves an average 1.17$\times$ (up to 1.69$\times$) end-to-end speedup across three representative servers and three video diffusion models, and reduces communication volume by 12.54--23.05\,\%. Moreover, we achieve an average 1.53--1.76$\times$ bandwidth improvements over NCCL.

\end{abstract}

%% file: sec/1_intro.tex
\section{Introduction}
\label{sec:intro}

Video generation has emerged as one of the most prominent generative AI applications \cite{seedance,luma,Veo,make_a_video,firely,mochi}. Among existing architectures, diffusion transformers (DiTs) have become the dominant backbone for video generation \cite{hacohen2024ltxvideorealtimevideolatent,kong2024hunyuanvideo,seedance2026seedance20advancingvideo,MiniMaxH32026,wan2025}. A DiT encodes a video as a sequence of visual tokens, each representing a small spatiotemporal region. Starting from random noise, the model iteratively predicts and removes noise to progressively recover the target video.

However, video diffusion typically involves hundreds of thousands of tokens, making multi-GPU inference necessary. Among DiT operators, attention is a major computational bottleneck because it models pairwise interactions between tokens, incurring quadratic complexity with sequence length.
To scale execution across multiple GPUs, prior studies adopt sequence parallelism (SP) \cite{liu2024ringattention,ulysses}, which partitions the token sequence across high-bandwidth GPU ranks, with each rank computing attention for its local queries. This distribution requires ranks to exchange the key and value (KV) data needed by remote queries at every attention layer. Although these high-bandwidth interconnects are widely used in generative AI research, their high cost means that commodity GPU clusters without such interconnects still constitute a substantial fraction of deployed infrastructure \cite{Mobius,ecoserve,Frugal}. On these platforms, limited inter-GPU bandwidth makes communication a central constraint on inference latency.

One promising direction for reducing the KV blocks required by each GPU is sparsification. 
However, existing CPU--GPU or GPU--GPU KV-block reduction techniques primarily focus on reducing data movement by predicting important KV blocks under the default token distribution policy \cite{ALISA,ECHO,OPKV,dsv}. However, queries with largely disjoint KV dependencies may produce a nearly dense union of required blocks on the same GPU, even when each individual attention mask is sparse. For example, when HunyuanVideo is partitioned across 8 GPUs, only 68.7\,\% of required KV blocks are accessed by more than 10\,\% of Q blocks, dropping to 39.8\,\% at 30\,\%. Thus, much of the communication serves only a small fraction of local Q blocks.
Beyond the KV demand induced by sparse attention kernels, the communication mechanism itself also determines the achievable reduction. The relay hops carry more data than the GPU actual need and reduce communication by 67.2\,\% (as shown in \Cref{fig:ring-saving}). A naive communication strategy may retain and forward unnecessary data during transfer. Moreover, default NCCL communication \cite{nccl} competes with attention computation for GPU execution resources, which can only achieve 43.8--61.1\,\% of the theoretical bandwidth.
To summarize, these inefficiencies degrade the communication benefits of sparsity and leave avoidable overhead on the critical path.

These observations impose three requirements on a communication efficient design. First, query placement should reduce the aggregate KV communication of each GPU, rather than only preserving per-query sparsity. However, placement must be determined before the exact layer-specific sparse masks become available, making it impractical to optimize for each realized mask independently. The key opportunity is that video attention exhibits persistent structural locality across layers: queries corresponding to similar spatial regions across frames tend to access overlapping KV regions \cite{sparse_vdit,xi2025sparse,adaspa}.
A good placement should therefore exploit this sparse pattern to co-locate queries with similar KV demands, reducing remote KV block movements. 

Second, the communication mechanism must preserve the savings created by sparse placement and sparse attention. Transferring KV blocks through relay GPUs that do not consume those blocks would introduce additional communication. Thus, the communication path should follow the actual producer--consumer dependency: each KV block should be delivered only from its owner to the consumer.

Third, reducing communication volume alone is insufficient to reduce end-to-end latency. Current distributed systems on commodity GPUs leverages GPU execution resources for both communication and computation. If communication contends with attention for shared resources, bandwidth will degrade and communication still remains exposed on the critical path.
Therefore, the runtime must allow communication to progress independently of attention.

\begin{figure}[t]
  \centering
  \includegraphics[width=\linewidth]{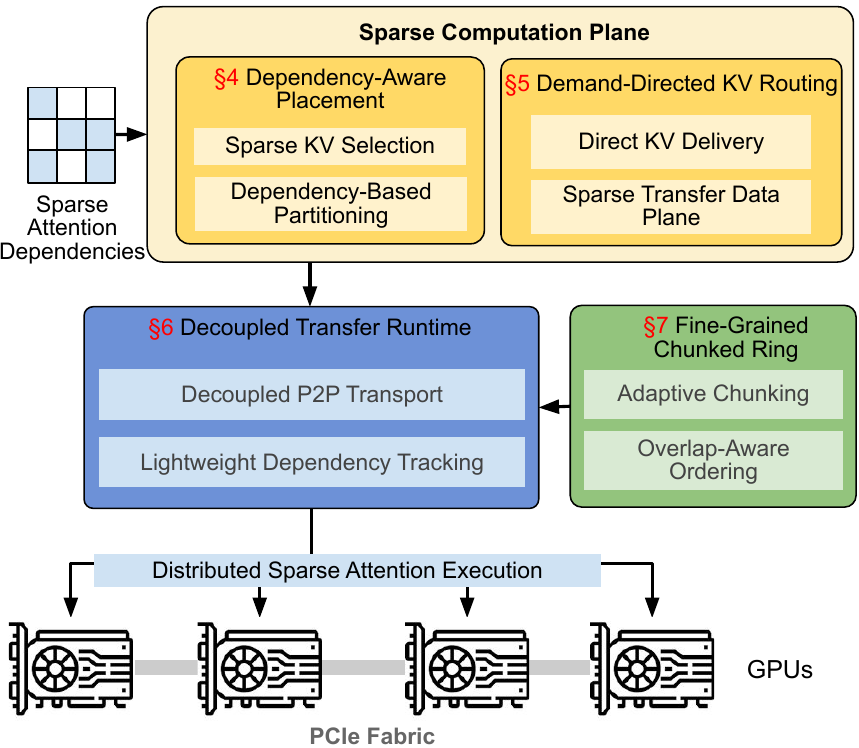}
  \caption{System overview.  Dependency-aware placement and direct routing
  reduce KV movement; the Decoupled Runtime and fine-grained chunking coordinate
  transfer progress and attention execution.}
  \label{fig:overview}
\end{figure}

Figure~\ref{fig:overview} shows how \name realizes these requirements. To reduce the aggregate KV demand of each GPU before layer-specific masks are known, \textit{Dependency-Aware Placement} exploits persistent cross-frame attention structure to colocate queries with overlapping KV dependencies (\S\ref{sec:reorder}). Once each layer constructs its sparse mask, \textit{Demand-Directed KV Routing} converts the resulting receiver requests into direct owner-to-consumer KV transfers, preventing sparsity-oblivious relay traffic (\S\ref{sec:cross-ring}). To translate the reduced traffic into latency savings, the \textit{Decoupled Transfer Runtime} moves KV data through copy engines and tracks fine-grained readiness independently of sparse-attention execution (\S\ref{sec:ce}). Finally, the \textit{Fine-Grained Chunked Ring} consumes KV blocks as soon as their dependencies are satisfied to minimize the pipeline drain at the end of the pipeline (\S\ref{sec:tail-chunk}).

We integrate \name into three video diffusion models: LTX-Video, LTX-2, and HunyuanVideo. 
Across three representative PCIe-connected platforms, \name achieves an average 1.38--1.5$\times$ attention speedup while reducing communication by 12.54--23.05\,\%. Its Decoupled Runtime further improves communication bandwidth by 1.53--1.76$\times$ over NCCL. Overall, \name delivers an average 1.17$\times$ end-to-end speedup over the best existing communication policy. This paper makes the following contributions:

\begin{itemize}[noitemsep,leftmargin=*]
\item An analysis of sparsity underutilization that characterizes bandwidth waste in distributed sparse attention.

\item A communication design that optimizes communication savings exposed by sparse attention.

\item A Decoupled Runtime that enables asynchronous data movement, together with a chunk-wise scheduler that advances attention as partial payloads become available to reduce pipeline drain.

\end{itemize}

%% file: sec/2_back.tex
\section{Background and Related Work}
\label{sec:background}

In this section, we introduce video diffusion models, sparse attention, and parallelism schemes in serving these models. 

\subsection{Video Diffusion Transformer Model}
Diffusion models have become the dominant paradigm for video generation due to their ability to generate high-quality visual content \cite{Tora,wan2025,hacohen2024ltxvideorealtimevideolatent,kong2024hunyuanvideo,MiniMaxH32026,PanoDiT}. Given a user prompt, these models typically start from randomly initialized noise in a compressed video representation, known as the latent state, and iteratively recover the target sample through denoising. At each denoising step, the model predicts the noise component in the current latent and removes it to progressively refine the generated video.

Modern video diffusion models typically implement this iterative process with diffusion transformers \cite{wan2025,hacohen2024ltxvideorealtimevideolatent,kong2024hunyuanvideo,MiniMaxH32026}. The core component is the denoising process, which consists of a sequence of transformer blocks and updates the latent for a fixed number of steps. Within these transformer blocks, the dominant operation is attention. Due to the quadratic compute complexity of attention kernels, it usually dominates the computation time of the diffusion transformer, especially for long videos which have hundreds of thousands of tokens.
Therefore, some researchers propose to exploit the inherent sparsity to accelerate video generation without hurting the output video's quality \cite{zhang2025spargeattention,sparse_vdit,adaspa,xi2025sparse}. 

\subsection{Sparse Attention in Video Diffusion Models}
The sparse attention implementation is based on the insight that not every computation is equally important \cite{zhang2025spargeattention,xi2025sparse,adaspa,Radial_Attention,svg2,sparse_vdit}. In diffusion models, the attention computes
\begin{equation}
  \mathbf{O} =
  \operatorname{softmax}\!\left(
  \frac{\mathbf{Q}\mathbf{K}^{\mathsf{T}}}{\sqrt{d}}
  \right)\mathbf{V},
  \label{eq:attention}
\end{equation}
where $Q$, $K$, and $V$ correspond to three token sequences, and $d$ denotes their hidden dimension. These studies observe that, for each token in $Q$, the output of $\operatorname{softmax}\left(
\frac{\mathbf{Q}\mathbf{K}^{\mathsf{T}}}{\sqrt{d}}
\right)$ is highly skewed: a small fraction of the $K$ tokens accounts for the majority of the softmax mass. Therefore, sparse attention prunes unnecessary attention scores after softmax and uses the resulting sparse attention matrix to compute the weighted aggregation over $V$.
While this technique has been shown to be effective for single-GPU generation, its scalability remains underexplored.

\subsection{Context and Sequence Parallelism}

\begin{figure}
\begin{subfigure}[t]{0.98\linewidth}
\centering
\includegraphics[width=1\linewidth]{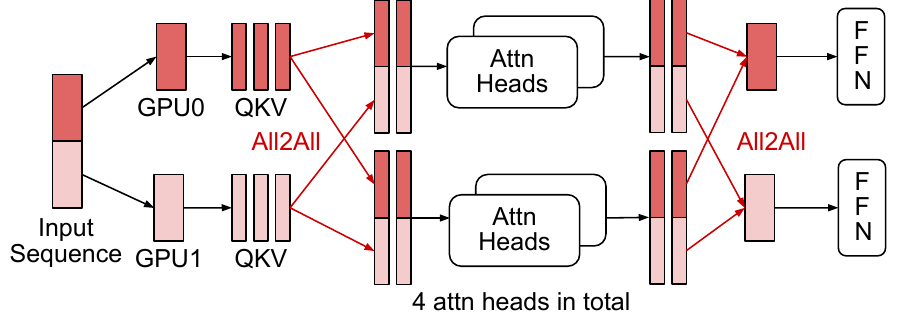}
\caption{Illustration of DeepSpeed Ulysses parallel algorithm. \label{fig:ulysses}}
\end{subfigure}
\begin{subfigure}[t]{0.98\linewidth} 
\centering
\includegraphics[width=1\linewidth]{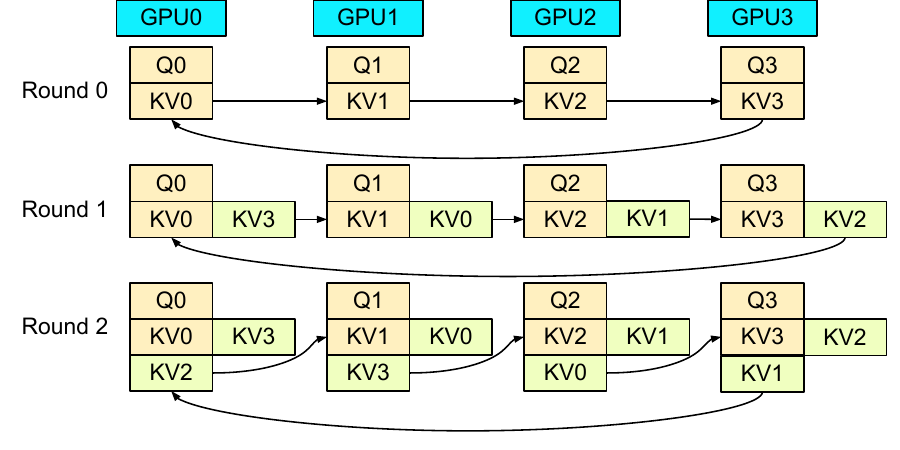}
\caption{Illustration of Ring attention parallel algorithm. \label{fig:ring}}
\end{subfigure}
\caption{Workflow of two sequence-parallel techniques.}
\label{fig:sp}
\end{figure}

To reduce diffusion latency, prior studies exploit parallel execution strategies that partition diffusion inference across multiple GPUs \cite{luo2025accelerating,fang2025pipefusion,distrifusion,yang2025scalefusion}. However, these techniques either rely on stale or compressed remote data, or require NVLink-level bandwidth to remain efficient. Video diffusion optimization on PCIe-connected servers remains underexplored, despite such GPUs constituting a substantial fraction of cloud resources \cite{ecoserve,Mobius,Frugal}.
Under their limited bandwidth, sequence parallelism provides a suitable mechanism because it preserves precision with less communication volume.

Sequence-parallel systems distribute a long activation sequence across devices to reduce per-device memory consumption. Two widely used sequence-parallel algorithms are Ulysses \cite{ulysses} and Ring Attention \cite{liu2024ringattention}. \Cref{fig:sp} illustrates the workflows of these two algorithms. Ulysses exchanges sequence and head partitions through all-to-all communication. As shown in \Cref{fig:ulysses}, each GPU holds only a portion of the full sequence. During attention, each GPU processes only a subset of the attention heads; therefore, Ulysses sends only the corresponding KV data to other GPUs. This design keeps the communication volume fixed as the number of GPUs increases. However, it cannot effectively overlap communication with computation.  
Ring Attention is another sequence-parallel method that rotates KV shards across GPUs. As illustrated in \Cref{fig:ring}, in the first round, each GPU sends its local KV shard to the next GPU. In subsequent rounds, each GPU forwards the KV shard received in the previous round to the next GPU. While KV shards are being transferred, each GPU computes online softmax between its local queries and the currently available KV shard, thereby overlapping communication with computation. Although Ring Attention enables such overlap, its communication volume grows with the number of GPUs, making communication increasingly difficult to hide behind computation.
Therefore, addressing the communication bottleneck on commodity GPUs is critical for efficient video diffusion inference.
We next analyze the detailed challenges and motivations for communication-efficient sparse attention.

%% file: sec/3_motiv.tex
\section{Motivation}
\label{sec:motivation}

Sparse attention exposes a fine-grained dependency graph in which each Query block consumes only a small subset of the available KV blocks. A distributed runtime realizes the corresponding savings only if it preserves this sparsity end-to-end. We observe that the sparsity surviving during communication is discounted by three independent factors: the union of demands aggregated on each rank (placement), the extra hops each selected block traverses (routing), and the fraction of the remaining transfer left exposed on the critical path (scheduling). Existing context-parallel runtimes address none of them: they pack KV blocks regardless of the dependency graph and relay them through intermediate GPUs. We next quantify each factor in turn (\S\ref{sec:sparse-shifts-bottleneck}--\S\ref{sec:pipeline-drain})

\subsection{Bottleneck Shifts to Communication}
\label{sec:sparse-shifts-bottleneck}

Our first experiment quantifies how much kernel-level sparsity survives at the communication layer.  For LTX, LTX2, and HunyuanVideo, we measure the
communication reduction in sparse attention. Since LTX-2 has two stages, we note them as L2 s1 and L2 s2.
We then partition each workload across eight GPUs and assume an optimistic exchange that every rank receives only the KV blocks required by its local queries.
This setup provides an upper bound on the communication savings available without changing query placement.

\begin{figure}[t]
  \centering
  \includegraphics[width=\linewidth]{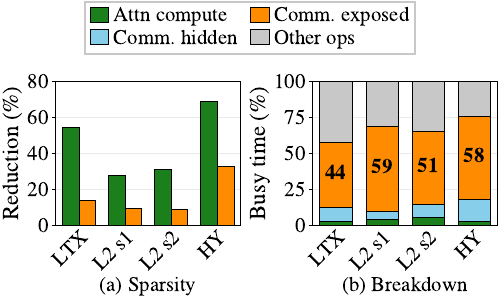}
  \caption{Compute sparsity outpaces communication sparsity.
  (a) Computation and communication savings enabled by sparsity.
  (b) Time breakdown of dense and sparse attention.}
  \label{fig:comm-overlap}
\end{figure}

Figure~\ref{fig:comm-overlap}(a) reveals an imbalance between computation and communication: sparse attention eliminates up to 71.3\,\% of computation, yet reduces communication by only 18.9\,\%. This imbalance matters because reducing computation also shrinks the opportunity to hide communication through overlap.
Consequently, on eight PCIe-connected L40S GPUs, communication becomes the dominant component of sparse-attention execution, as shown in Figure~\ref{fig:comm-overlap}(b). Kernel-level acceleration therefore shifts the bottleneck to the distributed data path rather than eliminating it.

\begin{takeaway}{box:comm-bottleneck}{}
The sparsification reduces the computation, yet leaves communication as the new bottleneck.
\end{takeaway}

\subsection{Bandwidth Waste by Naive Partition}
\label{sec:rank-union}

\begin{figure}[t]
  \centering
  \includegraphics[width=0.9\linewidth]{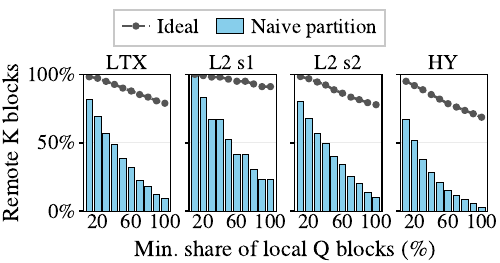}
  \caption{Remote K is inefficient with naive partition.}
  \label{fig:ideal-saving}
\end{figure}

The first source of communication overhead arises from naive sequence partitioning poorly aggregating sparse KV dependencies.We measure how efficiently remote K blocks are shared by local Q blocks. For each remote K block received by a rank, we compute the fraction of local Q blocks that access it. The ideal case assumes that all Q blocks requesting the same K block are co-located on a single rank. Figure~\ref{fig:ideal-saving} quantifies the resulting gap. A point \((x,y)\) means that \(y\,\%\) of the remote K blocks are accessed by at least \(x\,\%\) of the local Q blocks on the receiving rank.

With naive continuous sequence partitioning, remote K blocks are often required by only a small subset of local queries. The main reason is that naive partitioning scatters Q blocks with shared dependencies across ranks, causing the same K blocks to be repeatedly transferred to multiple ranks and wasting communication bandwidth.
Consequently, preserving bandwidth efficiency requires co-locating queries with highly overlapping KV dependencies.
For Ulysses, sparsity cannot be directly exploited because each GPU processes the full sequence for its assigned attention heads, whereas Ring Attention partitions query blocks, enabling communication to be reduced by transferring only the required KV blocks.
Therefore, this work focuses on Ring Attention.

\begin{takeaway}{box:naive-split}{}
Communication reduction is governed by dependency overlap within each rank, not by mask density alone.
\end{takeaway}

\subsection{Ring Forwarding Amplifies Sparse Traffic}
\label{sec:ring-amplification}

\begin{figure}[t]
  \centering
  \includegraphics[width=0.9\linewidth]{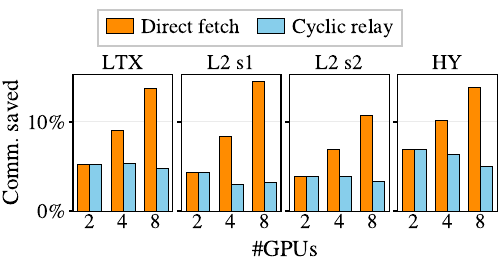}
  \caption{Ring forwarding amplifies sparse communication across GPU numbers. }
  \label{fig:ring-saving}
\end{figure}

Minimizing each rank's request volume is necessary but insufficient, because the routing topology can still cause sparse payloads to traverse unnecessary links. A rank's requested KV volume captures what it ultimately consumes, but should not contain payloads that will be forwarded to the next hop. Figure~\ref{fig:ring-saving} separates these two quantities by comparing consumer demand with the traffic generated by conventional Ring Attention.
With two GPUs, they act the same because each payload traverses only a single hop.
As the ring grows, however, more bytes are transferred than requested.

This divergence arises from cyclic forwarding. Each intermediate rank relays blocks requested by downstream consumers regardless of whether it uses those blocks locally. Consequently, a block selected by a single receiver may consume bandwidth on multiple links, and this amplification grows with ring size. 
Therefore, preserving sparse demand at the transport layer requires receiver-directed communication rather than unconditional forwarding. 

\begin{takeaway}{box:redundant-ring}{}
Conventional relay routing introduces redundant forward-hop transfers, wasting inter-GPU bandwidth.
\end{takeaway}

\subsection{Pipeline Drain Exposes the Critical Tail}
\label{sec:pipeline-drain}

The third source of unexpected communication overhead is an execution schedule that fails to exploit all available computation for communication hiding. We measure the communication and computation times of three models and compare Ring Attention's actual overlap against an ideal schedule that can utilize all independent attention work. Figure~\ref{fig:overlap-gap} reveals a persistent gap on PCIe: although Ring Attention overlaps a substantial fraction of each transfer with computation, a non-negligible portion of communication remains exposed on the critical path.

The gap arises from pipeline drain and coarse transfer granularity. When a PCIe transfer outlasts the sparse kernel scheduled immediately before it, the uncovered tail stalls the dependent consumer. After the payload arrives, the final attention kernel executes during pipeline drain, even though part of this computation could have started earlier as smaller payload fragments became available. Thus, the system has more computation to hide communication, but the execution schedule cannot effectively utilize it. 

\begin{takeaway}{box:overlap}{}
Pipeline drain leaves computation exposed on the critical path, reducing communication--computation overlap.
\end{takeaway}

\textbf{Summary.}
These observations motivate the design of \name. Guided by the insight in \Cref{box:comm-bottleneck}, \name targets communication efficiency for sparse video diffusion models on commodity GPUs. First, \name employs Dependency-Aware Placement (\S\ref{sec:reorder}) to reduce rank-level demand amplification by co-locating queries with overlapping KV dependencies, following the insight in \Cref{box:naive-split}. Second, to realize the resulting communication savings and address the redundancy identified in \Cref{box:redundant-ring}, \name introduces Demand-Directed KV Routing (\S\ref{sec:cross-ring}) to replace multi-hop forwarding with receiver-directed transfers. Building on these two designs, \name further incorporates a Decoupled Transfer Runtime (\S\ref{sec:ce}) to materialize sparse transfers efficiently with higher bandwidth. Finally, \name employs a Fine-Grained Chunked Ring (\S\ref{sec:tail-chunk}) to reduce exposed communication latency by improving communication--computation overlap, following the insight in \Cref{box:overlap}.

%% file: sec/4_selectkv.tex
\section{Dependency-Aware Placement}
\label{sec:reorder}

Section~\ref{sec:rank-union} shows that naively aggregating demanded KV blocks at rank granularity can introduce substantial redundant communication. \name addresses this inefficiency through \emph{Dependency-Aware Placement}, which partitions video tokens according to their sparse attention dependencies. We next describe its underlying principle and design.

\begin{figure}[t]
\begin{minipage}[b]{0.45\linewidth}
    \centering
    \includegraphics[width=\linewidth]{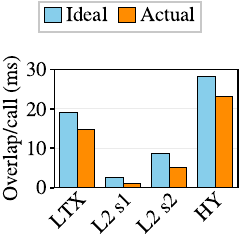}
    \caption{Communication--computation overlap.}
    \label{fig:overlap-gap}
\end{minipage}
\hfill
\begin{minipage}[b]{0.45\linewidth}
    \centering
    \includegraphics[width=\linewidth]{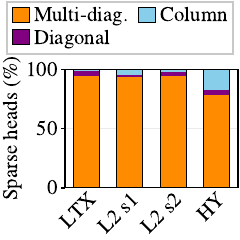}
    \caption{Distribution of sparse attention patterns.}
    \label{fig:sparse-pattern}
\end{minipage}
\end{figure}

\subsection{Design Overview}
We formulate sparse exchange around the block dependency graph exposed by the attention backend.  Let $M_{h,i,j}\in\{0,1\}$ indicate whether query block $i$ in head $h$ consumes KV block $j$.  Let $Q_r$ denote the query blocks assigned to rank $r$, and let $P(j)$ identify the rank that owns KV block $j$.  Rank $r$'s request to owner $p$ for head $h$ is
\begin{equation}
  D_{r,h,p}=\left\{j\mid P(j)=p\land
  \exists i\in Q_r:M_{h,i,j}=1\right\}.
  \label{eq:receiver-demand}
\end{equation}
The set union in Equation~\ref{eq:receiver-demand} is precisely the
rank-level aggregation measured in Section~\ref{sec:rank-union}.  For a KV
block size of $b$ tokens, a head dimension of $d$, and $w$ bytes per element,
the payload received by rank $r$ is
\begin{equation}
  V_r(P)=2bdw\sum_h\sum_{p\ne r}|D_{r,h,p}|,
  \label{eq:sparse-exchange-volume}
\end{equation}
where the factor of two accounts for K and V.
Equation~\ref{eq:sparse-exchange-volume} exposes the two decisions made in this
section: selective routing removes blocks outside $D_{r,h,p}$, while
dependency-aware placement changes $Q_r$ and $P$ to reduce the demand unions themselves.

\subsection{Sparse KV Selection}
\label{sec:selective-kv}
An intuitive way to reduce communication is to transfer only the KV blocks required by each GPU. We next describe how \name enables such selective communication.

\textbf{Obtaining dependencies before payload exchange}
A sparse data path must determine $D_{r,h,p}$ before moving the full KV
payload.  This ordering is difficult for token-level methods whose sparse mask
is available only after attention begins.  Prior work predicts the mask with a
lightweight model with shrunk token dimensions~\cite{dsv}, but
maintaining prediction quality requires online fine-tuning and adds work to the inference path.

\name instead targets block-sparse video DiT kernels that materialize their masks from pooled query and key tiles~\cite{zhang2025spargeattention,hu2026dfsattn,gu2026blade,NEURIPS_vsa}.
Before launching the attention kernel, each rank exchanges pooled key tiles rather than the full K tensor. These summaries are tens to hundreds of times smaller than the original keys, and their communication can overlap with value projection. Compared with transferring the full K tensor, this overhead is low. Each receiver then combines the remote key summaries with its local query summaries to construct $M$ and materialize Equation~\ref{eq:receiver-demand} entirely on the GPU. Thus, request generation directly reuses information already produced by the sparse attention kernel, without requiring a predictor or additional approximation.

\subsection{Dependency-Based Partitioning}
\label{sec:pattern-aware}

\begin{figure}[t]
  \centering
  \includegraphics[width=\linewidth]{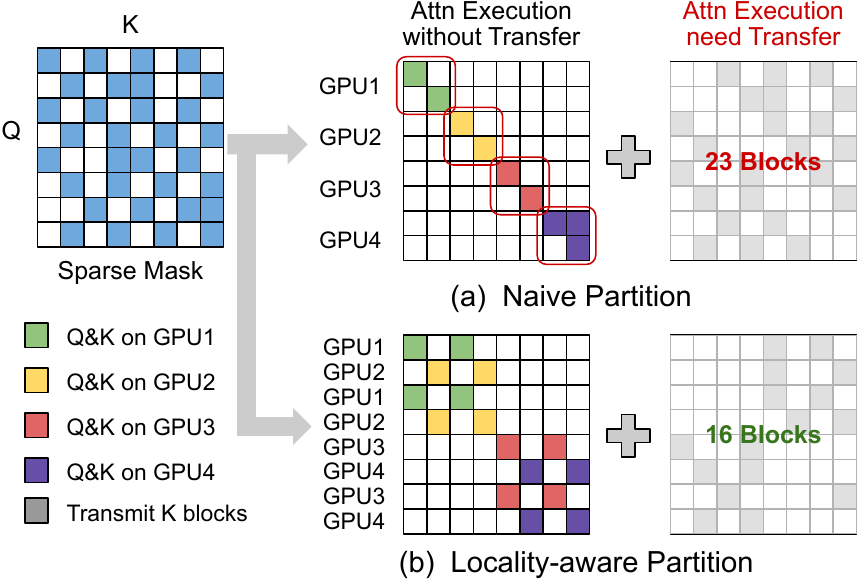}
  \caption{Dependency-aware placement reduces rank-level union amplification.
  (a) Contiguous sequence partitioning transfers more KV blocks because
  co-located queries have dissimilar dependencies.  (b) Grouping queries with
  overlapping dependencies reduces the transfer.}
  \label{fig:reorder}
\end{figure}

Selective routing eliminates blocks outside each $D_{r,h,p}$, but it cannot
control the size of the set union in Equation~\ref{eq:receiver-demand}.  Figure~\ref{fig:reorder} provides a concrete example of why naive partitioning misses opportunities for communication reduction. A contiguous partition assigns the first two query blocks to the same GPU even though their KV demands overlap little, causing the four ranks to transfer 23 blocks in total. Reassigning queries with similar dependencies to the same rank increases local reuse and reduces the transfer volume to 16 blocks without modifying the attention mask. Placement therefore complements request filtering: filtering removes unselected blocks, whereas placement reduces the number of distinct selected blocks required by each rank.

The runtime cannot optimize placement for every mask.  A denoising
step contains thousands of attention heads, where each head has its own sparse mask.
Moreover, the sequence must be partitioned before the sparse masks become available. To guide this partitioning decision, we analyze the distribution of sparse attention patterns in video diffusion models. As shown in Figure~\ref{fig:sparse-pattern}, video attention primarily exhibits three forms: diagonal dependencies capture intra-frame neighborhoods, multi-diagonal dependencies connect corresponding regions across frames, and column dependencies identify globally important KV blocks. Multi-diagonal dependencies dominate across all three models. Because their periodicity is determined by the frame size, video geometry provides a stable placement signal—the diagonal interval—even when the exact sparse mask varies across attention layers.

Therefore, \name uses an interval-periodic layout.  It divides each frame into $G$ contiguous spatial intervals and assigns the same interval to the same rank in every frame for $B$ blocks:
\begin{equation}
  P_{\mathrm{periodic}}(t,s)=
  \left\lfloor\frac{Gs}{B}\right\rfloor.
  \label{eq:periodic-placement}
\end{equation}
where $t$ and $s$ are its frame and block index.
The ownership pattern repeats at every frame boundary.  Queries that attend to the same spatial region across frames consequently share an owner, which turns repeated multi-diagonal dependencies into local accesses or smaller owner-specific unions. 

Consider a head in which every query at spatial position $s$ attends to the same neighborhood $[s-\delta,s+\delta]$ across each of the $T$ frames. A balanced rank owns $B/G$ spatial positions. Among all sets of this cardinality, a single contiguous interval minimizes the size of its $\delta$-neighborhood: away from frame boundaries, the resulting neighborhood spans $B/G+2\delta$ positions. Replicating this interval across frames therefore bounds the remote demand by at most $2\delta T$ halo blocks, up to sequence-boundary effects. Any fragmented assignment introduces additional boundaries and cannot yield a smaller demand union.
The periodic layout is therefore optimal under the multi-diagonal demand
model.

\name realizes Equation~\ref{eq:periodic-placement} as a deterministic permutation from logical block coordinates to $(\mathit{rank}, \mathit{offset})$ pairs. It computes this permutation once for a given model geometry and SP degree, applies it consistently to the hidden states and all position-dependent metadata, and reuses it across attention layers and denoising steps. The permutation operates at the granularity of the kernel's native KV blocks without altering block membership, thereby preserving the mathematical correctness. Once the distributed region is complete, the inverse permutation restores the model's original token order. As a result, the critical path requires neither mask-dependent repartitioning nor host-side placement decisions.

%% file: sec/5_ring.tex
\section{Demand-Directed KV Routing}
\label{sec:cross-ring}

\begin{figure}[t]
  \centering
  \includegraphics[width=\linewidth]{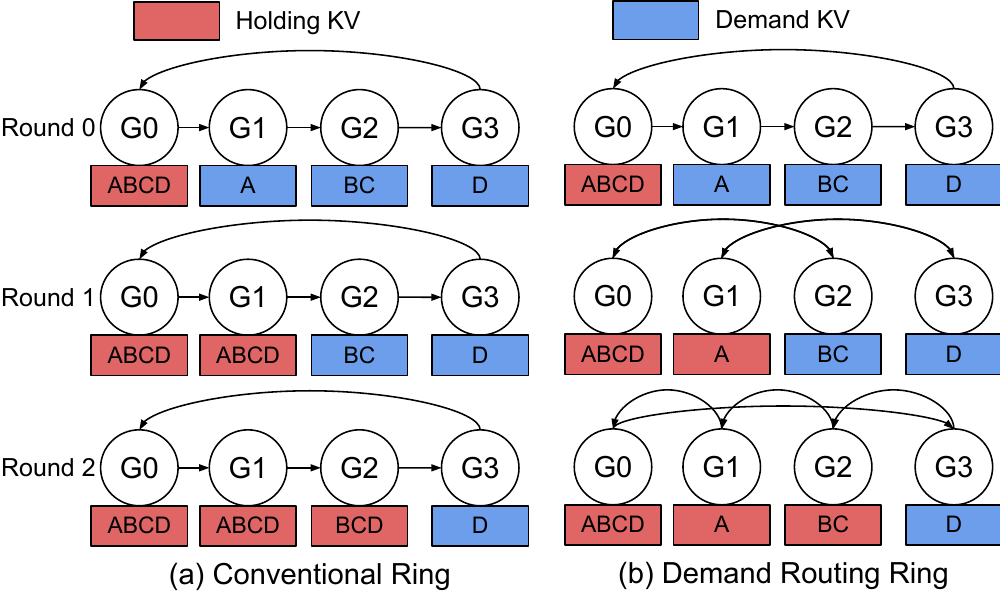}
  \caption{(a) Conventional Ring Attention carries unnecessary tokens due to the demand of the next hop. (b) Demand-Directed KV Routing only transfers necessary tokens. }
  \label{fig:cross-ring}
\end{figure}

Although the previous section reduces aggregate KV demand, these savings cannot be fully realized under the relay-based routing of conventional sequence parallelism. Therefore, we redesign the Ring data plane to materialize the communication savings exposed by sparse demand.

\subsection{Direct KV Delivery}

Figure~\ref{fig:cross-ring}(a) illustrates the limitation of conventional Ring Attention from GPU$_0$'s perspective.  Suppose GPU$_0$ owns four KV blocks, $A$--$D$, while GPU$_1$, GPU$_2$, and GPU$_3$ require only $A$, $BC$, and $D$, respectively.  In the first round, GPU$_0$ must send the complete shard $ABCD$ to GPU$_1$. Although GPU$_1$ consumes only $A$, it must retain and forward $BCD$ because downstream GPUs may need them.  This way, selective KV reduces the destination demand, but the fixed forwarding path prevents these savings from translating directly into communication reduction.

\name replaces state circulation with \emph{demand-routed exchange}. Rather than forwarding a shard hop by hop, a producer communicates directly with the consumer that requires each selected block. 
As shown in Figure~\ref{fig:cross-ring}(b), round~0 communicates with consumers at ring distance one, round~1 with consumers at distance two, and so on. In round $k$, each rank therefore communicates directly with the consumer at distance $k$, while transmitting only the KV blocks selected by that consumer.

This schedule eliminates relay-induced redundancy: a block crosses the interconnect only when it is required by the destination GPU, rather than because it happens to lie on the forwarding path of another consumer.  The resulting communication volume therefore tracks the actual sparse KV demand instead of the size of the dense shard.

\subsection{Sparse Transfer Data Plane}
Demand routing exposes a sparse producer--consumer graph, but naively materializing it can reintroduce control overhead because different heads request different numbers of remote KV blocks.  \name avoids this overhead by preserving demand at head granularity and establishing the complete communication plan once per attention call.

\textbf{Head-preserving demand.}
For each remote peer, \name represents demand as a Boolean matrix over $(\text{head}$, $\text{KV-block})$.  We do not union requests across heads: different heads often select complementary blocks, causing the union to approach the complete shard even when individual heads remain sparse. Instead, \name packs and communicates KV blocks independently per head, preserving the sparsity exposed by the attention mask.

\textbf{Metadata-free exchange.}
Each rank computes its demand for all $G-1$ remote peers and exchanges these masks in one batched point-to-point operation.  
After this step, producer $p$'s \texttt{Provide} mask is identical to consumer $c$'s \texttt{Demand} mask.  Both endpoints can therefore enumerate the same $(\text{head},\text{block})$ sequence locally and in the same order. \name packs blocks in row-major order, making each head a contiguous range in the packed buffer.  Metadata, such as message lengths, per-head offsets, and receive ranges, is thus derived locally, eliminating per-transfer descriptors. 

%% file: sec/6_ce.tex
\section{Decoupled Transfer Runtime}
\label{sec:ce}

Reducing KV traffic alone does not eliminate the communication bottleneck. The remaining transfers must still progress concurrently with the sparse attention kernels that consume them. Moreover, existing communication libraries impose ordering and buffer-management constraints that are more conservative than required by the sparse producer--consumer dependency graph. To address these inefficiencies, \name uses DMA as its communication route, decoupling data movement from GPU kernel execution.

\subsection{Runtime Overview}
\name treats KV exchange not as an opaque communication operation, but as a lightweight Decoupled Runtime whose data movement and synchronization are explicitly controlled by the sparse execution plan. The runtime exposes five primitives: (1) \textsc{Export} establishes reusable peer-visible slots; (2) \textsc{Publish} marks newly packed contents as ready; (3) \textsc{Fetch} moves a receiver-derived byte range through the asynchronous peer-copy path; (4) \textsc{Release} returns a slot to its producer as soon as the consumer no longer requires its contents;  (5)\textsc{Fence} provides a coarse safety boundary between successive attention invocations.

\begin{figure}[t]
  \centering
  \includegraphics[width=1\linewidth]{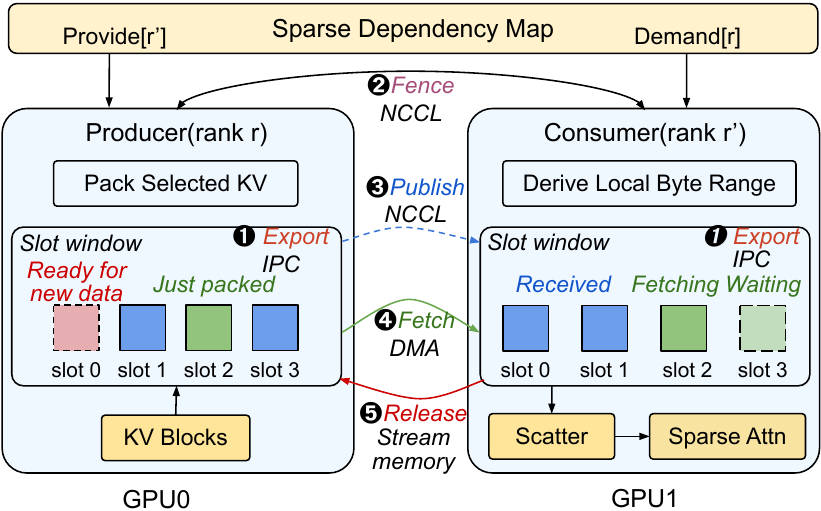}
  \caption{Illustration of Decoupled Transfer Runtime.}
  \label{fig:ce}
\end{figure}

Figure~\ref{fig:ce} illustrates the resulting data path. Before model execution, all GPUs expose their slot windows through \textsc{Export} (\circled{1}) using CUDA IPC memory handles. These slot windows remain unchanged unless the model or execution configuration changes. Data destined for a given peer is stored in a single slot rather than split across slots, and therefore, a consumer reads only one slot from each remote GPU for each hop.
At the beginning of each attention operation, \name uses an NCCL barrier as the \textsc{Fence} (\circled{2}) primitive to ensure that all GPUs have completed communication from the previous attention. Once all ranks are ready, producers begin preparing the data required by their consumers. For each producer--consumer pair, the producer packs the corresponding payload into the designated slot (slot 2 in the figure). After packing completes, the producer invokes \textsc{Publish} (\circled{3}), implemented as a 1-byte NCCL message, to notify the consumer that the slot is ready. Because \textsc{Publish} is issued on the same CUDA stream as the packing operation, CUDA guarantees that the payload is fully prepared before notification.
The consumer then invokes \textsc{Fetch} (\circled{4}), which uses a CUDA peer copy through DMA to retrieve the payload from the remote GPU. 
After the transfer completes, the consumer writes a 4-byte integer to a page shared across ranks. The producer polls this state to determine whether the corresponding slot can be reclaimed and reused for subsequent data preparation; this operation serves as \textsc{Release}~(\circled{5}).

\subsection{Decoupled P2P Transport}
\label{sec:compute-decoupled}

Kernel-driven transport can introduce unexpected resource contention because computation and communication compete for the same SM resources. 
Moreover, existing communication libraries employ general producer--consumer protocols to manage buffer lifetimes, incurring synchronization overhead that is unnecessary for Ring Attention, where both producers and consumers already know the communication metadata. 
Therefore, we use raw asynchronous peer copies, which leverage DMA engines for inter-GPU data movement. 

\label{sec:receiver-address}

To start such a memory copy process, the consumer should input the target address or size information.
To avoid carrying additional descriptors during transfer, all ranks export slots using the same geometry and packing convention. Consequently, a consumer can reconstruct the producer's packing order and derive the byte range corresponding to any requested head or block subrange using only locally available state. \textsc{Fetch} directly specifies the slot index for each provider--consumer pair. Once the consumer receives the \textsc{Publish} signal, it can immediately pull data from the corresponding GPU.

The transferred representation packs only the selected KV blocks contiguously. Upon arrival, a lightweight scatter step places these blocks into the addressing layout expected by the sparse attention kernel.

\subsection{Lightweight Dependency Tracking}
\label{sec:prefetch-pipeline}

A hop-level global synchronization trivially guarantees execution order, but forces all ranks to progress at the pace of the slowest participant. Alternatively, allocating a fresh destination buffer for every transfer avoids reuse hazards, but causes memory consumption to scale with the number of outstanding transfers. \name instead manages a bounded window of reusable slots with explicit producer--consumer dependencies. All dependencies are expressed through \textsc{Publish} and \textsc{Release}, rather than reconstructed within the transport layer. Because \name already knows the dependency structure of each hop, the communication primitive does not need to infer it again.
Therefore, the producer can advance independently of consumer completion as long as free communication slots remain. Consumer acknowledgments are processed asynchronously and gate the producer's next pack. Each rank maintains multiple slots, allowing subsequent data to be prefetched even when a slow consumer delays the release of an earlier slot.

In contrast, \textsc{Fence} provides a coarser-grained guarantee. At attention call boundaries, it ensures that all peer ranks have completed the previous attention before the slot-window state is reset or reused by the next one. This asymmetric synchronization keeps the common-case path local to each producer--consumer pair while preserving a coarse mechanism for correctness across attention calls. The global synchronization occurs only once per attention call, matching the synchronization frequency of existing SP strategies.

%% file: sec/7_chunk.tex
\section{Fine-Grained Chunked Ring}
\label{sec:tail-chunk}

Even after reducing KV traffic and decoupling transport from computation, communication can still dominate sparse attention on bandwidth-limited PCIe fabrics. In a Ring Attention pipeline, when transfer time exceeds computation time, the next attention block cannot begin immediately, leaving part of the communication exposed on the critical path. The final remote block introduces an additional \emph{pipeline drain}: although earlier computation can overlap with communication, the final computation stage has no subsequent transfer to overlap with. Consequently, an asynchronous Ring pipeline may still incur a substantial exposed tail from both communication and computation.

\subsection{Adaptive Chunking}
\label{sec:adaptive-chunking}

\name reduces this tail by changing transfer granularity only where finer-grained execution can shorten the critical path.  Instead of uniformly dividing every remote block, \name uses an \emph{Adaptive Chunking} policy that selects a configuration-specific chunk count for different positions in the Ring pipeline.

\begin{figure}[t]
  \centering
  \includegraphics[width=\linewidth]{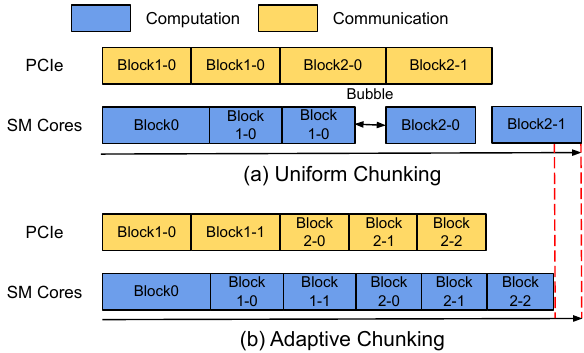}
  \caption{(a) Conventional Ring Attention with uniform chunk sizes. (b) Adaptive chunk size tuned by phases to achieve better overlap. }
  \label{fig:tail-chunk}
\end{figure}

The local block remains intact because it requires no communication. Remote blocks, however, may use different chunk counts depending on their pipeline position and communication-to-computation ratio. Earlier blocks already overlap with subsequent transfers, so excessively fine chunking mainly increases launch and scheduling overhead. The final remote block, in contrast, directly determines the pipeline drain and therefore benefits most from finer granularity: computation on its first chunk can begin as soon as that chunk arrives, rather than waiting for the entire block. As shown in Figure~\ref{fig:tail-chunk}, the first remote blocks retain the default chunk configuration, while the final remote block is divided into finer-grained chunks.
This allows computation on the tail block to start earlier and progressively overlap with the arrival of its remaining chunks, reducing the exposed drain latency. Accordingly, \name independently tunes the chunk counts for the steady-state and drain phases, avoiding unnecessarily fine chunking where communication is already well hidden.

\subsection{Overlap-Aware Ordering}
\label{sec:chunk-ordering}

Chunking exposes an additional scheduling opportunity. \name partitions attention by heads, but these heads are heterogeneous: different chunks contain different numbers of selected KV blocks and therefore incur different communication and computation costs. Executing chunks in their original order can leave an unnecessarily large computation tail after the final communication completes.
Therefore, \name orders chunks according to their contribution to the critical path. 
In the communication-dominated regime, computation for earlier chunks can overlap with transfers of later chunks, whereas the computation of the final scheduled chunk necessarily remains exposed at the end of the pipeline. 
Thus, \name schedules the chunk with the highest sparsity last, minimizing the non-overlapped computation tail.

%% file: sec/8_eval.tex
\section{Evaluation}
\label{sec:evaluation}

\begin{figure}[t]
  \centering
  \includegraphics[width=\linewidth]{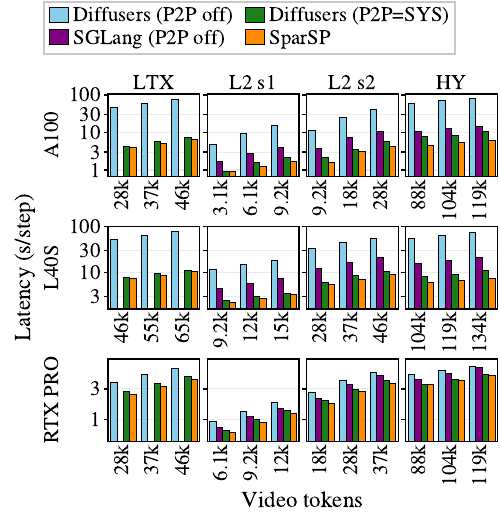}
  \caption{End-to-end execution latency.\protect\footnotemark}
  \label{fig:e2e}
\end{figure}

\footnotetext{Note: SGLang does not support LTX-Video.\label{ft:sglang}}

\subsection{Setup}
 
\textbf{Platforms:} We evaluate \name on three representative PCIe-connected GPU platforms. 
(a) The \textbf{first platform} is a server with 8 A100 SXM4 GPUs connected via PCIe 4.0 $\times$ 16. Although A100 SXM4 GPUs support high-bandwidth interconnects, some cloud providers still offer PCIe-only configurations without NVLink connectivity for cost-sensitive users. The host is a dual-socket AMD EPYC system, with four GPUs attached to each socket; consequently, cross-group GPU traffic traverses the inter-socket fabric. Each GPU has 40 GB of VRAM. (b) The \textbf{second platform} contains 8 L40S GPUs connected via PCIe 4.0 $\times$ 16. Its GPU topology is identical to that of the first platform, and each GPU has 48 GB of VRAM. (c) The \textbf{third platform} contains 4 RTX 6000 PRO GPUs connected via PCIe 5.0. All GPUs communicate through PCIe Host Bridges within the same NUMA node, and each GPU has 96 GB of VRAM.

\textbf{Benchmarks:}We evaluate \name using three representative text-to-video diffusion models: LTX-Video \cite{hacohen2024ltxvideorealtimevideolatent}, LTX-2 \cite{hacohen2026ltx2efficientjointaudiovisual}, and HunyuanVideo \cite{kong2024hunyuanvideo}. For LTX-Video and LTX-2, we use a default frame resolution of 768 $\times$ 1536, while HunyuanVideo uses 768 $\times$ 1280. We evaluate \name across different sequence lengths by varying the number of video frames. LTX-2 consists of two stages, denoted as s1 and s2 in our evaluation: the first stage operates at a lower resolution to establish the overall layout, while the second stage refines visual details at a higher resolution. Due to memory limitations, we additionally use fp8 precision for LTX-Video and LTX-2 on A100 and L40S GPUs. We adopt weight-only quantization while keeping all activations in bf16 precision, ensuring that the observed communication benefits do not arise from reduced activation precision.

We use SpargeAttn \cite{zhang2025spargeattention} as the sparse attention kernel and adopt its CDF-based sparsification mode. For each query token, CDF mode retains the key blocks with the highest attention scores until their cumulative attention mass reaches the specified \textit{CDF threshold}. We set the default CDF thresholds to 0.85, 0.9, and 0.85 for LTX-Video, LTX-2, and HunyuanVideo, respectively. We sample prompts from VidProM \cite{VidProM}. For each method, we execute 200 denoising steps and report the average per-step latency.

\textbf{Baselines:} We implement \name{} in 2.5K lines of Python and C++/CUDA code on top of Diffusers \cite{von-platen-etal-2022-diffusers}. We compare \name{} against the following baselines. (1) \textbf{Diffusers} \cite{von-platen-etal-2022-diffusers}: We port SpargeAttn to Diffusers to ensure a fair comparison and use Ring Attention as the sequence-parallel communication backend. (2) \textbf{SGLang} \cite{zheng2024sglangefficientexecutionstructured}: We also evaluate against SGLang using the same SpargeAttn sparse-attention backend for consistency. On commodity GPU platforms, SGLang adopts Ulysses as its default sequence-parallel strategy. Since SGLang does not support LTX-Video, we compare \name{} against SGLang only on the other two models. To isolate the benefits of sequence parallelism, we disable all other forms of parallelism in SGLang.

\begin{figure}[t]
  \centering
  \includegraphics[width=\linewidth]{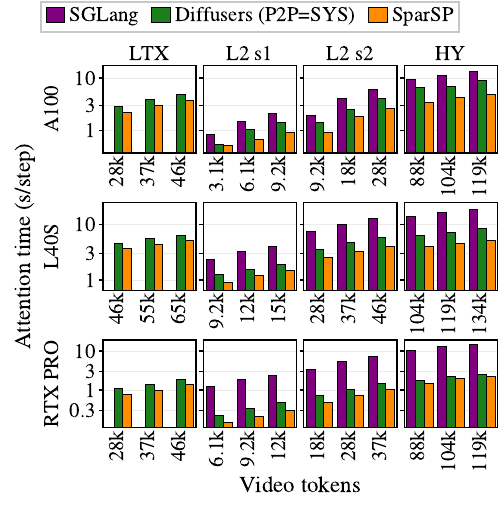}
  \caption{Attention execution latency in one step.\footref{ft:sglang}}
  \label{fig:e2e-attn}
\end{figure}

\subsection{End-to-End Performance}
\label{sec:current-results}
We first evaluate latency reduction across different sequence lengths on all three servers. For Diffusers, the default NCCL configuration provides limited bandwidth. Explicitly setting the P2P level to SYS substantially improves Ring Attention's point-to-point bandwidth \footnote{NCCL uses a conservative default cutoff for GPU P2P communication, typically allowing P2P only for paths up to the PXB level. On our platform, GPU pairs reported as NODE by nvidia-smi topo traverse PCIe host bridges and are therefore beyond this default cutoff. As a result, NCCL disables direct P2P transport and falls back to host-backed shared-memory communication. Setting NCCL\_P2P\_LEVEL=SYS relaxes the cutoff and enables direct GPU peer access for these pairs, substantially improving bandwidth.}, but degrades Ulysses' All-to-All performance.  We therefore report Diffusers under both the default and P2P=SYS configurations. Figure \ref{fig:e2e} reports the average per-step latency of the three models across the three platforms. \name consistently achieves the lowest latency, delivering an average speedup of 1.17$\times$ and 1.88$\times$ over Diffusers (P2P=SYS) and SGLang, respectively. Notably, Diffusers can outperform SGLang once its P2P bandwidth is improved, further highlighting the dominance of communication overhead.

\begin{figure}[t]
  \centering
  \includegraphics[width=\linewidth]{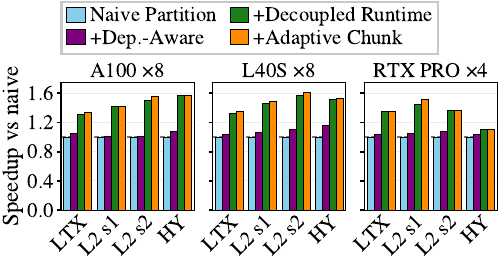}
  \caption{Optimization breakdown.}
  \label{fig:breakdown}
\end{figure}

The gain is governed by each configuration's communication-to-computation ratio, and therefore varies predictably rather than arbitrarily. A100 pairs strong compute and memory bandwidth with cross-socket PCIe, giving the highest ratio and up to 1.69$\times$ (1.3$\times$ on average); L40S shares the same topology but with slower compute, and RTX 6000 PRO combines PCIe 5.0 with a modest compute increase, leaving Ring Attention able to hide most communication and yielding only 1.07$\times$. 
The same trend holds within a single model: LTX-2's half-resolution stage s1 has far less attention work than its full-resolution stage s2 on identical hardware, and gains accordingly less.

We further measure attention latency within each denoising step. Figure \ref{fig:e2e-attn} reports the attention latency across the evaluated configurations. Compared with the end-to-end results, \name achieves larger improvements at the attention level, confirming that attention communication is the primary source of its performance gains. 
\name outperforms Diffusers by 1.38--1.5$\times$ on average across the three platforms. 
Notably, \name achieves an average 2.7$\times$ speedup over SGLang in attention latency, much higher than its end-to-end speedup. 
This gap arises because SGLang has already optimized other operators, while attention and its associated communication dominate the remaining latency.

\subsection{Ablation Study}
\label{sec:current-ablation}

\textbf{Benefits Breakdown:} We further evaluate the contribution of each design in \name. To isolate the benefit of sequence partitioning, we use Demand-Directed KV Routing for all configurations in this experiment and quantify the benefit of Demand-Directed KV Routing separately later. Figure \ref{fig:breakdown} shows the speedup contributed by each component. 
We use naive partitioning, which is adopted by prior sparsity-aware communication systems \cite{dsv}, as the baseline and report speedup over it. Across all three platforms, the Decoupled Runtime provides the largest improvement, ranging from 1.2--1.4$\times$, because the default communication path limits transfer bandwidth. HunyuanVideo benefits the most from Dependency-Aware partitioning, achieving a 1.04--1.09$\times$ speedup due to its high sparsity. In contrast, tail chunking provides no benefit for HunyuanVideo because its smaller number of attention heads leaves limited room for chunking, and the additional kernel-launch overhead outweighs the overlap benefits.

\begin{figure}[t]
  \centering
  \includegraphics[width=\linewidth]{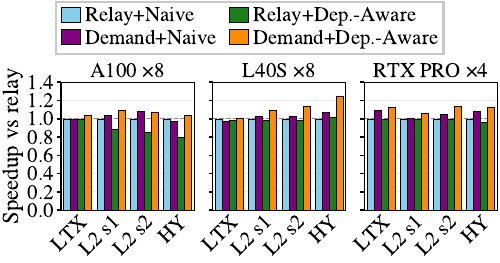}
  \caption{Benefits from Demand-Directed KV Routing.}
  \label{fig:ring-poloicy}
\end{figure}

\begin{figure}[t]
  \centering
  \includegraphics[width=\linewidth]{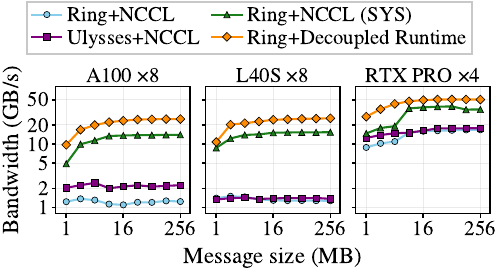}
  \caption{Bandwidth of NCCL and \name.}
  \label{fig:bandwidth}
\end{figure}

\textbf{Demand-Directed KV Routing Benefits:} We next evaluate the performance benefit of our Demand-Directed KV Routing policy. We measure the attention latency of all three models across the three platforms, with and without Demand-Directed KV Routing, under both naive partitioning and our Dependency-Aware partitioning policy. We normalize attention latency to naive partitioning with conventional Ring Attention. Figure \ref{fig:ring-poloicy} shows that adopting Dependency-Aware partitioning alone can even degrade performance because it may increase relay traffic under conventional Ring routing. Demand-Directed KV Routing alone can degrade performance on A100 and L40S servers because it introduces additional cross-NUMA communication. It therefore needs to be combined with Dependency-Aware partitioning to provide sufficient communication reduction for a net performance gain. In contrast, on the single-NUMA RTX 6000 PRO platform, Demand-Directed KV Routing consistently reduces latency.

\textbf{Bandwidth:} We further measure the effective bandwidth of different communication methods across the three platforms. Figure \ref{fig:bandwidth} reports the achieved bandwidth under varying message sizes. The A100 and L40S platforms exhibit relatively low bandwidth due to cross-NUMA communication, whereas our Decoupled Runtime consistently achieves the highest bandwidth across all platforms. Setting the P2P connection level to SYS substantially improves Ring Attention bandwidth. Nevertheless, our Decoupled Runtime still outperforms this configuration by 1.53--1.76$\times$.

\subsection{Sensitivity Study}

\begin{figure}[t]
  \centering
  \includegraphics[width=\linewidth]{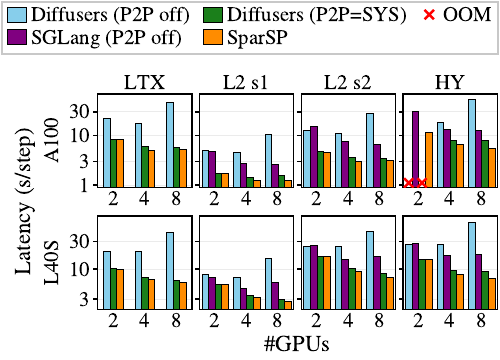}
  \caption{End-to-end latency over different GPU numbers.\footref{ft:sglang}}
  \label{fig:e2e-scaling}
\end{figure}

\begin{figure}[t]
  \centering
  \includegraphics[width=\linewidth]{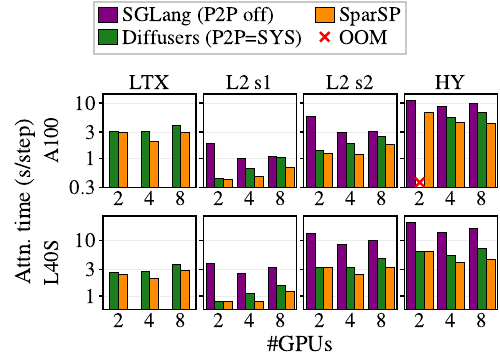}
  \caption{Attention latency over different GPU numbers.\footref{ft:sglang}
  }
  \label{fig:attn-scaling}
\end{figure}


\textbf{Scalability:} To evaluate the scalability of \name, we measure end-to-end latency on 2, 4, and 8 A100 and L40S GPUs. Figure \ref{fig:e2e-scaling} and Figure \ref{fig:attn-scaling} report the end-to-end and attention latency, respectively. 
On 2 A100 GPUs, Diffusers fails due to OOM.
As the GPU count increases, \name achieves lower latency, although the scaling benefit gradually diminishes. 
This is because the computation per GPU is halved, while the total communication of Ring Attention increases, limiting the gains from additional parallelism. As communication becomes more dominant at larger GPU counts, \name achieves increasingly larger speedup over Diffusers. With 2 GPUs, computation hides most of the communication, leaving limited optimization opportunity. As the GPU count increases, the end-to-end speedup of \name over Diffusers grows from 1.01$\times$ to 1.21$\times$.

\begin{figure}[t]
  \centering
  \includegraphics[width=0.9\linewidth]{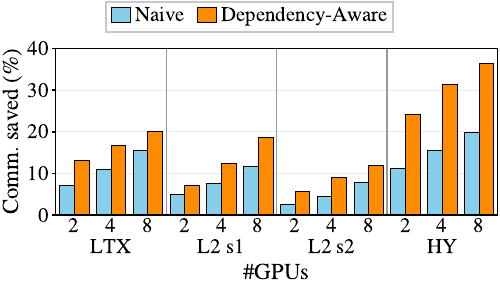}
  \caption{Communication reduction over GPU numbers}
  \label{fig:comm_scaling}
\end{figure}

\textbf{Communication Reduction:} We further analyze communication reduction under different GPU counts. Figure \ref{fig:comm_scaling} compares the two partitioning policies. As the number of GPUs increases, communication reduction also increases because each rank aggregates fewer query dependencies, reducing the union of demanded KV blocks, as discussed in \S\ref{sec:rank-union}. Across all models and GPU counts, Dependency-Aware Partition consistently achieves greater communication savings than naive partitioning. The largest reduction is observed on HunyuanVideo, indicating that its attention patterns exhibit stronger cross-frame dependencies and provide more opportunities for dependency-aware placement.

\begin{figure}[t]
  \centering
  \includegraphics[width=0.9\linewidth]{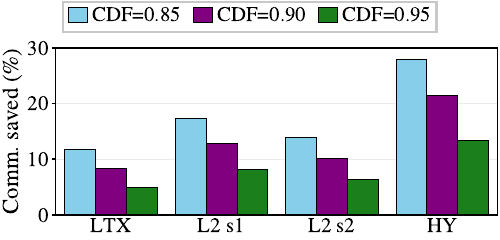}
  \caption{Communication reduction vs. sparsity.}
  \label{fig:cdf_comm}
\end{figure}

\textbf{Sparsity:} We further evaluate communication reduction under different sparsity levels. We vary the CDF threshold of SpargeAttn from 0.85 to 0.95 and measure the communication savings achieved by our Dependency-Aware partition policy on 8 GPUs. Figure \ref{fig:cdf_comm} shows that communication reduction decreases as the CDF threshold becomes more restrictive. Nevertheless, even at a threshold of 0.95, \name achieves an average communication reduction of 12.54\,\% across the three models. At a threshold of 0.85, the average reduction increases to 23.05\,\%. These results demonstrate that \name remains effective even under relatively dense attention patterns, highlighting its robustness across different sparsity levels.

%% file: sec/9_con.tex
\section{Related works}

\textbf{Video Acceleration:} Prior studies have proposed various techniques to accelerate video diffusion generation, with caching and sparsification being two of the most common approaches. FlexCache \cite{sun2024flexcacheflexibleapproximatecache} exploits approximate caching to improve video-serving throughput, while TeaCache \cite{teaCache}, FasterCache \cite{lv2025fastercache}, and AdaCache \cite{zihao2026adacache} leverage similarities across adjacent diffusion steps to avoid redundant computation. In parallel, SpargeAttn \cite{zhang2025spargeattention}, Sparse VideoGen \cite{xi2025sparse}, and Sparse-vDiT \cite{sparse_vdit} exploit non-uniform attention importance and design efficient sparse-attention kernels to accelerate video generation. Although effective, these techniques primarily target single-GPU computation rather than inter-GPU communication, making them complementary to our system.

\textbf{Distributed Sparse Computations:} Another line of research focuses on improving sparse computation in distributed systems. SpComm3D \cite{abubaker2024spcomm3dframeworkenablingsparse} exploits static sparsity to replace sparsity-agnostic bulk collectives with selective point-to-point communication. However, this design is not directly applicable to video diffusion models, whose sparse masks are dynamic and unavailable before execution. PecSched \cite{zhang2025pecschedpreemptiveefficientcluster} considers GPU topology and avoids unnecessary communication over distant GPU links. db-SP \cite{chen2026dbsp} and FVAttn \cite{liu2026fvattnadaptivesparseattention} primarily improve GPU load balance rather than communication efficiency. Therefore, these techniques are complementary to and composable with \name.

\section{Conclusion}
\label{sec:conclusion}
We have presented \name, a distributed runtime that exploits sparsity to enable efficient video diffusion communication on PCIe-connected commodity GPUs. Our key insight is that computational sparsity does not directly translate into communication savings; realizing such savings requires jointly optimizing data placement, routing, transport, and communication--computation overlap. \name integrates dependency-guided exchange, demand-routed communication, a Decoupled Transfer Runtime, and chunk-wise overlap scheduling under a unified sparse-demand representation. On representative commercial text-to-video diffusion models, \name achieves an average $1.3\times$ end-to-end speedup on a representative server over the baseline. 